\documentclass[epj]{webofc}
\usepackage[utf8]{inputenc}
\usepackage[varg]{txfonts}   % Web of Conferences font
\usepackage{booktabs}
\usepackage{xcolor}
\definecolor{darkred}{rgb}{0.4,0.0,0.0}
\definecolor{darkgreen}{rgb}{0.0,0.4,0.0}
\definecolor{darkblue}{rgb}{0.0,0.0,0.4}
\usepackage[bookmarks,linktocpage,colorlinks,
    linkcolor = darkred,
    urlcolor  = darkblue,
    citecolor = darkgreen]{hyperref}
\usepackage{subfigure}
\usepackage{graphicx}
\usepackage{amsmath}
\usepackage{amssymb}

\wocname{EPJ Web of Conferences}
\woctitle{Lattice2017}

\newcommand{\be}{\begin{equation}}
\newcommand{\ee}{\end{equation}}
\newcommand{\bea}{\begin{eqnarray}}
\newcommand{\eea}{\end{eqnarray}}
\newcommand{\bi}{\begin{itemize}}
\newcommand{\ei}{\end{itemize}}
\newcommand{\ben}{\begin{enumerate}}
\newcommand{\een}{\end{enumerate}}
\newcommand{\bt}{\begin{tabbing}}
\newcommand{\et}{\end{tabbing}}
\newcommand{\nn}{\nonumber}

\newcommand{\calO}{{\mathcal O}}

\newcommand{\pp}{{p^\prime}}
\newcommand{\bfp}{{\bf p}}
\newcommand{\bfpp}{{{\bf p}^\prime}}

\newcommand{\bfx}{{\bf x}}
\newcommand{\bfxp}{{{\bf x}^\prime}}
\newcommand{\bfxpp}{{{\bf x}^{\prime\prime}}}

\newcommand{\dt}{{\Delta x_4}}
\newcommand{\dtp}{{\Delta x_4^\prime}}
\newcommand{\fpzDP}{{f_{\{+,0\}}^{DP}}}
\newcommand{\fpDP}{{f_+^{DP}}}

\newcommand{\fzDP}{{f_0^{DP}}}
\newcommand{\fpzDpi}{{f_{\{+,0\}}^{D\pi}}}
\newcommand{\fpDpi}{{f_+}^{D\pi}}
\newcommand{\fzDpi}{{f_0}^{D\pi}}
\newcommand{\fpDK}{{f_+}^{DK}}
\newcommand{\fpzDK}{{f_{\{+,0\}}^{DK}}}

\newcommand{\fpDKpi}{{f_+}^{DK(\pi)}}
\newcommand{\fparaDP}{{f_v^{DP}}}
\newcommand{\fperpDP}{{f_p^{DP}}}
\newcommand{\fhqetDP}{{f_{\{v,p\}}^{DP}}}
\newcommand{\fparaDK}{{f_v^{DK}}}
\newcommand{\fperpDK}{{f_p^{DK}}}
\newcommand{\fhqetDK}{{f_{\{v,p\}}^{DK}}}

\newcommand{\fhqetDpi}{{f_{\{v,p\}}^{D\pi}}}
\newcommand{\tsrc}{{x_{4,\rm src}}}

\begin{document}
%%%%%%%%%%%%%%%%%%%%%%%%%%%%%%%%%%%%%%%%%%%%%%%%%%%%%%%%%%%%%%%%%%%%%%%%%%%%%
%
\selectlanguage{english}
%----------------------------------------------------------------------------
\title{%
$D$ meson semileptonic form factors in $N_f\!=\!3$ QCD
with M\"obius domain-wall quarks
}
%----------------------------------------------------------------------------
\author{%
\firstname{Takashi} \lastname{Kaneko}\inst{1,2} \fnsep\thanks{Speaker, \email{takashi.kaneko@kek.jp}} \and
\firstname{Brian} \lastname{Colquhoun}\inst{1} \and
\firstname{Hidenori}  \lastname{Fukaya}\inst{3} \and 
\firstname{Shoji}  \lastname{Hashimoto}\inst{1,2}
\\(JLQCD Collaboration)
}
%----------------------------------------------------------------------------
\institute{%
High Energy Accelerator Research Organization (KEK),
Ibaraki 305-0801, Japan 
\and
SOKENDAI (The Graduate University for Advanced Studies),
Ibaraki 305-0801, Japan
\and
Osaka University, 
Osaka 560-0043, Japan
}
%----------------------------------------------------------------------------
\abstract{%
  We present our calculation of $D\!\to\!\pi$ and $D\!\to\!K$ semileptonic
  form factors in $N_f\!=\!2+1$ lattice QCD.
  We simulate three lattice cutoffs $a^{-1}\!\simeq\!2.5$, 3.6 and 4.5~GeV
  with pion masses as low as 230~MeV.
  The M\"obius domain-wall action is employed for both light and charm quarks.
  We present our results for the vector and scalar form factors and discuss
  their dependence on the lattice spacing, light quark masses
  and momentum transfer.
}
%---------
\maketitle
%---------

%// introduction ---------------------------------------------------------------

\section{Introduction}\label{sec:intro}

The $D\!\to\!K\ell\nu$ and $\pi\ell\nu$ semileptonic decays
provide a precise determination of the
Cabibbo-Kabayashi-Maskawa (CKM) matrix elements $|V_{cs}|$ and $|V_{cd}|$,
respectively.
% through the $c\!\to\!s\ell\nu$ and $d\ell\nu$ transitions,
% respectively.
The hadronic matrix element for the $D\!\to\!P\ell\nu$ decay ($P\!=\!\pi,K$)
is described by the vector and scalar form factors $f_{\{+,0\}}^{DP}$ 
as 
\begin{align}
   \langle P(\pp) | V_\mu | D(p) \rangle 
   \ = \ 
   \left(
      p+\pp - \frac{M_D^2-M_P^2}{q^2}q
   \right)_\mu
   \fpDP(q^2)
   +
   \frac{M_D^2-M_P^2}{q^2} q_\mu\, 
   \fzDP(q^2),
   \label{eqn:intro:ME}
\end{align}
where 
$q^2=(p-p^\prime)^2$ is the momentum transfer.
Due to parity symmetry of QCD,
the semileptonic and leptonic decays are sensitive
to interactions with different Dirac structures:
in the Standard Model, for instance,
only the weak vector (axial) current contributes
to the semileptonic (leptonic) decays.
These two decay processes are therefore complementary probes
of new physics.
In contrast to the leptonic decays,
the hadronic uncertainty of the form factors limits
the accuracy of the determination of $|V_{cs(d)}|$
and new physics search through the semileptonic decays,
due to less precise lattice QCD calculations % of the hadronic inputs
and more precise experimental data~\cite{Kaneko:2017ysl}.
The JLQCD Collaboration launched 
an independent calculation of the $D$ meson semileptonic form factors
on fine lattices~\cite{Fukaya:2016fzs,Kaneko:2017sct}.
This article updates the status of this project.

%// simulation method ----------------------------------------------------------
\section{Simulation method}\label{sec:sim}

%// action, 1/a, mud, ms

We simulate $N_f\!=\!2+1$ QCD
using the tree-level improved Symanzik gauge action
and the M\"obius domain wall quark action~\cite{Brower:2012vk}.
Our choice of the sign function approximation and the kernel operator
in its four-dimensional effective action
largely reduces the computational cost
compared to our previous works with the overlap fermion,
and it enables us to simulate large lattice cutoffs
$a^{-1}\!\gtrsim\!2.5$~GeV~\cite{Kaneko:2013jla}
with good chiral symmetry.
At $a^{-1}\!\sim\!2.5$ and 3.6 GeV,
we take three values of degenerate up and down quark mass, $m_{ud}$,
and two values of strange quark mass $m_s$
close to its physical value $m_{s,\rm phys}$.
The range of the pion mass is
$300~\mbox{MeV}\!\lesssim\!M_\pi\!\lesssim\!500$~MeV.
For a better control of the continuum and chiral extrapolations, 
we extend these simulations
to a finer lattice at $a^{-1}\!\sim\!4.5$~GeV with $M_\pi\!\sim\!300$~MeV,
as well as
to a lighter pion mass $M_\pi\!\sim\!230$~MeV at $a^{-1}\!\sim\!2.5$~GeV.
Depending on $a^{-1}$ and $m_{ud}$, 
we employ a spatial lattice size $L$
satisfying a condition $M_\pi L \!\gtrsim\!4$
in order to control finite volume effects.
The statistics are 5,000 Molecular Dynamics time at each simulation point.
Simulation parameters are summarized in Table~\ref{tbl:sim:param}.
The main progress since the previous report~\cite{Kaneko:2017sct}
is that we have completed two simulations at our largest $a^{-1}$
and smallest $M_\pi$.

At the large cutoffs 2.5\,--\,4.5~GeV,
we can also safely employ the same domain-wall action for charm quarks.
The charm quark mass is fixed to the physical value determined from
the low-lying charmonium spectrum.
The renormalized mass in the $\overline{\rm MS}$ scheme 
$m_c(3~\mbox{GeV})\!=\!1.003(10)$\,GeV~\cite{Nakayama:2016atf}
is consistent with the present world average.
We also note that
chiral symmetry is preserved to good accuracy at these cutoffs.
The residual quark mass is $O(1~\mbox{MeV})$ at $a^{-1}\!\sim\!2.5$~GeV
and even smaller ($\lesssim\!0.2$~MeV) on finer lattices
with moderate sizes in the fifth dimension $\sim\!10$.

\begin{table}[b]
\centering
\small
% \begin{flushleft}
\caption{
  Simulation parameters. Quark masses are bare value in lattice units.
  The rightmost column shows the number of
  temporal locations of the meson source operator
  of correlation functions.
}
% \end{flushleft}
% \vspace{0mm}
\label{tbl:sim:param}
\begin{tabular}{l|llll|l}
   \hline 
   lattice parameters 
   & $m_{ud}$ & $m_s$ & $M_\pi$[MeV] & $M_K$[MeV] & $N_\tsrc$ 
   \\ \hline
   $\beta\!=\!4.17$, \ 
   $a^{-1}\!=\!2.453(4)$, \ 
   $32^3\!\times\!64\!\times\!12$
   & 0.0190 & 0.0400 & 499(1) & 618(1) & 2
   \\
   & 0.0120 & 0.0400 & 399(1) & 577(1) & 2
   \\
   & 0.0070 & 0.0400 & 309(1) & 547(1) & 4
   \\
   & 0.0035 & 0.0400 & 226(1) & 525(1) & 4
   \\ \cline{2-6}
   & 0.0190 & 0.0300 & 498(1) & 563(1) & 2
   \\
   & 0.0120 & 0.0300 & 397(1) & 518(1) & 2
   \\
   & 0.0070 & 0.0300 & 310(1) & 486(1) & 4
   \\ \hline
   $\beta\!=\!4.35$, \ 
   $a^{-1}\!=\!3.610(9)$, \ 
   $48^3\!\times\!96\!\times\!8$
   & 0.0120 & 0.0250 & 501(2) & 620(2) & 2
   \\
   & 0.0080 & 0.0250 & 408(2) & 582(2) & 2
   \\
   & 0.0042 & 0.0250 & 300(1) & 547(2) & 4
   \\ \cline{2-6}
   & 0.0120 & 0.0180 & 499(1) & 557(2) & 2
   \\
   & 0.0080 & 0.0180 & 408(2) & 516(2) & 2
   \\
   & 0.0042 & 0.0180 & 296(2) & 474(2) & 4
   \\ \hline
   $\beta\!=\!4.47$, \ 
   $a^{-1}\!=\!4.496(9)$, \ 
   $64^3\!\times\!128\!\times\!8$
   & 0.0030 & 0.0150 & 284(1) & 486(1) & 4
   \\ \hline
\end{tabular}
% \vspace{0mm}
\end{table}

%// correlation functions

We extract the $D\!\to\!P$ matrix element (\ref{eqn:intro:ME})
from the three-point function
\begin{align}
    C^{DP}_{V_\mu}
    (\bfp,\bfp^\prime;\dt,\dtp)
    \ = \ 
    \frac{1}{N_s^3\,N_\tsrc}
    \sum_{\tsrc}
    \sum_{\bfx,\bfxp,\bfxpp} 
    \langle 
       {\calO}_P(\bfxp,\tsrc+\dt+\dtp) 
    \hspace{20mm}
    \nn \\
    \hspace{40mm}
       \ \times \ 
       V_\mu(\bfxpp,\tsrc+\dt)
       {\calO}_D^{\dagger}(\bfx,\tsrc)
    \rangle
    e^{-i\bfpp\left( \bfxp - \bfxpp \right)}
    e^{-i\bfp\left( \bfxpp - \bfx \right)},
    \hspace{10mm}
    \label{eqn:sim:msn_corr:3pt}
\end{align}
where 
$N_s\!=\!L/a$,
and ${\calO}_{\{D,P\}}$ represents the meson interpolating field
with a Gaussian smearing.
The initial $D$ meson is at rest ($\bfp\!=\!0$),
and we simulate four values of the momentum transfer $q^2$
by taking the final light meson momenta with $|\bfpp|^2\!=\!0$, 1, 2, 3
in units of $(2\pi/L)^2$.

%// coordinates

In our measurement of $C^{DP}_{V_\mu}(\bfp,\bfp^\prime;\dt,\dtp)$,
we fix the total temporal separation $\dt\!+\!\dtp$
and vary the location of the vector current $\dt$.
We choose $\dt\!+\!\dtp\!=\!28a$ at $a^{-1}\!\sim\!2.5$~GeV
by inspecting the stability of the form factor results
against $\dt\!+\!\dtp$~\cite{Fukaya:2016fzs}.
The physical length of $\dt\!+\!\dtp$ is the same for the other two cutoffs.
In order to improve the statistical accuracy,
$C^{DP}_{V_\mu}$ is averaged over
the location of the meson source operator $(\bfx,\tsrc)$
as indicated in Eq.~(\ref{eqn:sim:msn_corr:3pt}).
We employ the volume source generated with $Z_2$ noise
to average over $\bfx$ at a given time-slice $\tsrc$.
This measurement is repeated over $N_\tsrc$ different values of $\tsrc$.
Table~\ref{tbl:sim:param} shows our choice of $N_\tsrc$.
The correlation of $C^{DP}_{V_\mu}$ among different $\tsrc$'s
is not large with the small values of $N_\tsrc\!=\!2$\,--\,4.
We observe about a factor of $\sqrt{N_\tsrc}$ improvement
in the statistical accuracy.

%// extraction of FFs

\begin{figure}[tb] 
  \centering
  \includegraphics[width=0.49\linewidth,clip]{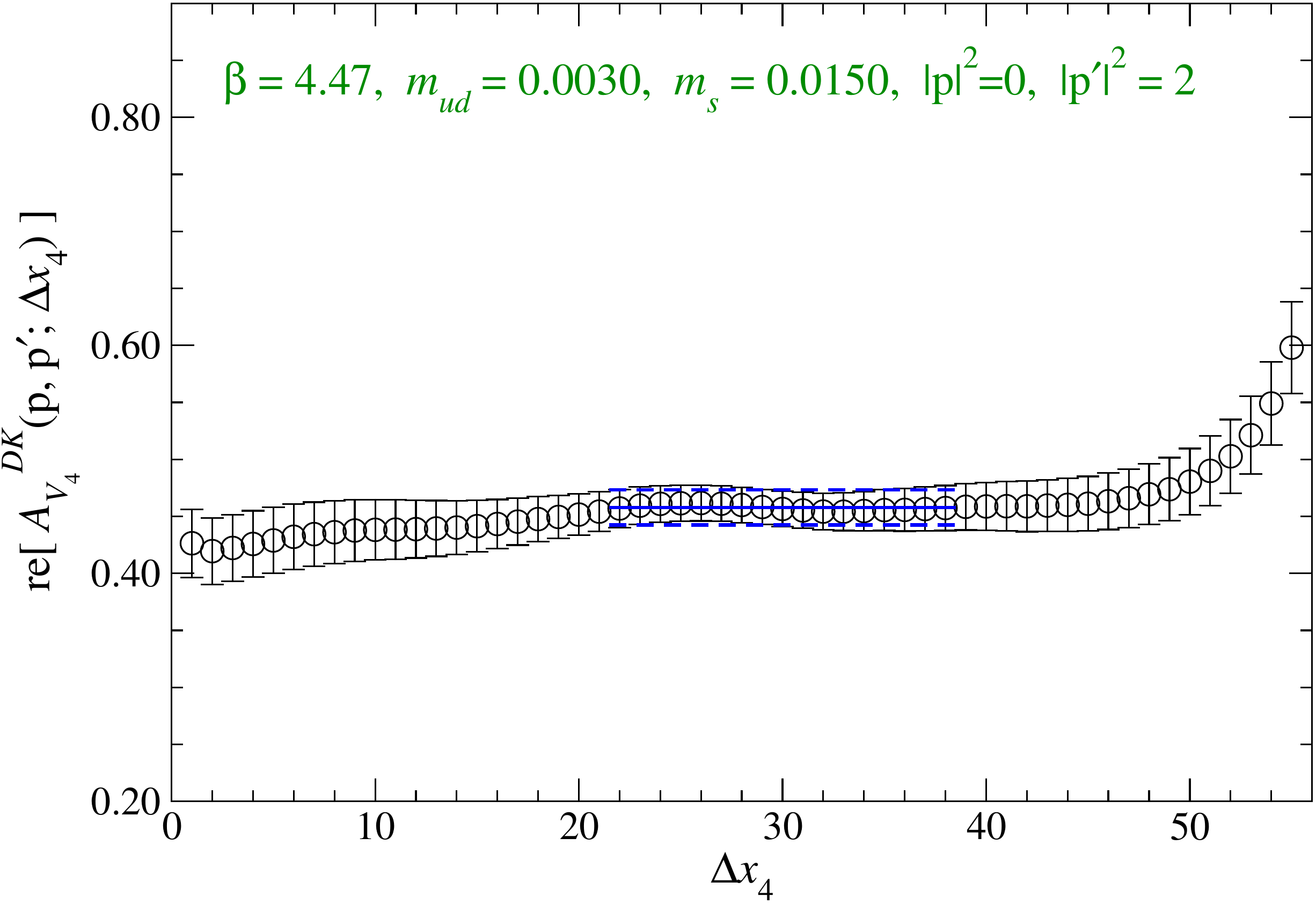}
  \hspace{1mm}
  \includegraphics[width=0.49\linewidth,clip]{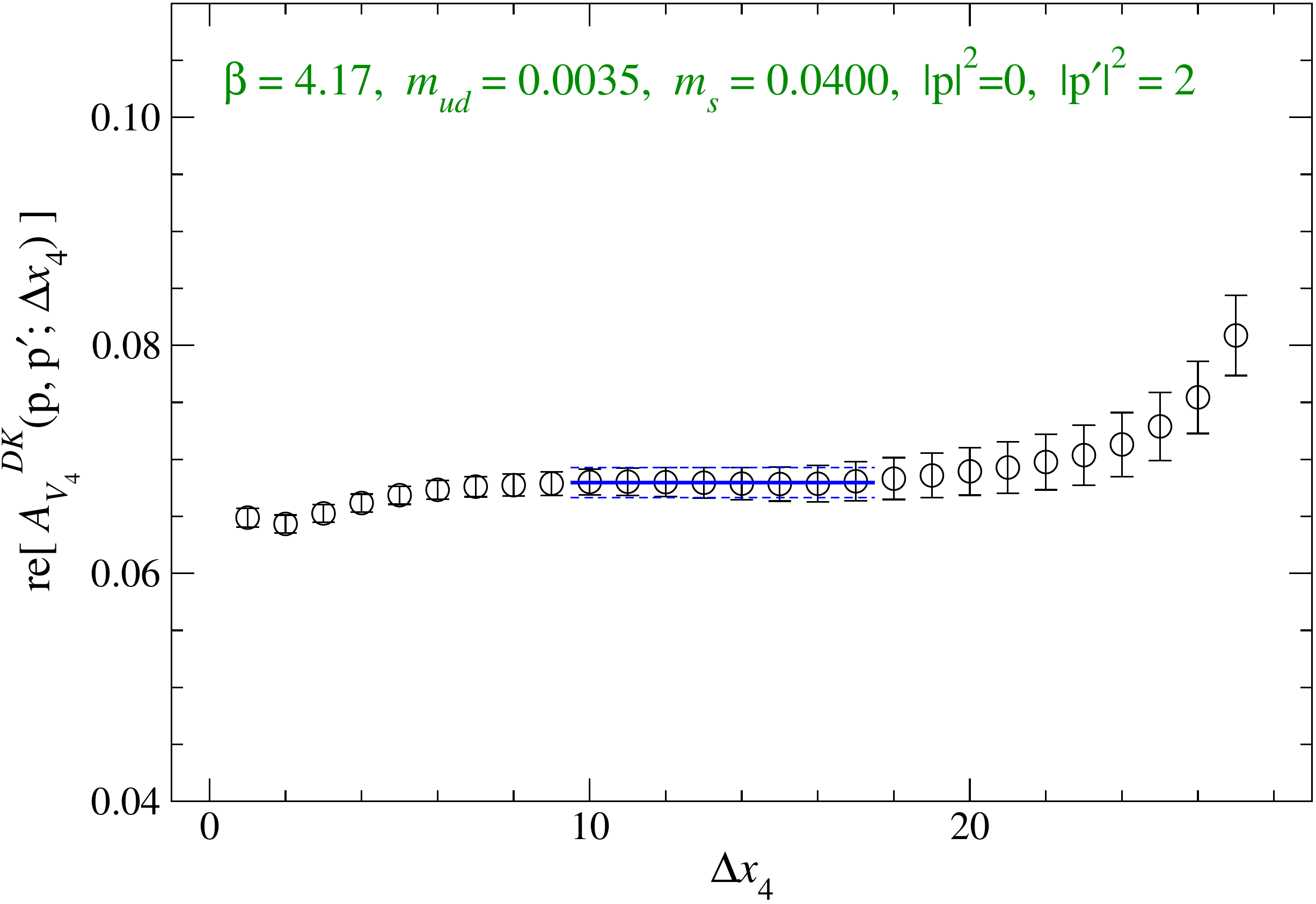}
  \caption{
    Effective value of amplitude $A_{V_4}^{DK}(\bfp,\bfpp)$
    for $D\!\to\!K$ decay as a function of $\dt$.
    The left panel shows data at the largest cutoff $a^{-1}\!\sim\!4.5$~GeV
    ($\beta\!=\!4.47$ and $M_\pi\!\simeq\!300$~MeV),
    whereas the right panel is for the smallest $M_\pi\!\simeq\!230$~MeV
    ($\beta\!=\!4.17$, $a^{-1}\!\sim\!2.5$~GeV).
    The $D$ meson is at rest, and the squared magnitude of the kaon momentum
    is $|\bfpp|^2\!=\!2$ in units of $(2\pi/L)^2$.
  }
  \label{fig:sim:amp}
  \vspace{-3mm}
\end{figure}

We also calculate two-point functions of $\pi$, $K$ and $D$ mesons
\begin{align}
    C^{Q}
    (\bfp;\dt)
    \ = \  
    \frac{1}{N_s^3\,N_\tsrc}
    \sum_{\tsrc}
    \sum_{\bfx,\bfxp} 
    \langle 
       {\calO}_Q(\bfxp,\tsrc+\dt) 
       {\calO}_Q^{\dagger}(\bfx,\tsrc)
    \rangle
    e^{-i\bfp\left( \bfxp - \bfx \right)}
    \hspace{2mm}
    (Q=\pi,K,D)
    \hspace{3mm}
    \label{eqn:sim:msn_corr:2pt}
\end{align}
in a similar way.
The amplitudes of these correlation functions are extracted from
the following fit in terms of $\dt$
\begin{align}
   C_{V_\mu}^{DP}(\bfp,\bfpp;\dt,\dtp)
   \ = \ 
   A_{V_\mu}^{DP}(\bfp,\bfpp) e^{-E_D(\bfp)\dt}e^{-E_P(\bfpp)\dtp}
   \hspace{5mm}
   (P\!=\!\pi,K),
   \label{eqn:sim:amp:3pt}
   \\ 
   C^Q(\bfp;\dt)
   \ = \ 
   B^Q(\bfp) e^{-E_Q(\bfp)\dt}
   \hspace{5mm}
   (Q\!=\!\pi,K,D),
   \hspace{20mm}
   \label{eqn:sim:amp:2pt}
\end{align}
where we estimate the meson energies $E_{\{\pi,K,D\}}$
from their rest masses and the dispersion relation in the continuum limit.
Figure~\ref{fig:sim:amp} shows examples of the effective value
of $A_{V_\mu}^{DP}(\bfp,\bfpp)$. 
We observe reasonably long plateaus and can also reliably determine
the amplitudes at the largest $a^{-1}$ and smallest $M_\pi$.

%// MEs and FFs

We evaluate the $D\!\to\!P$ matrix element as 
\begin{align}
   \langle P(\bfpp) | V_\mu | D(\bfp) \rangle
   \ = \ 
   2 Z_V
   \sqrt{
      \frac{ E_D(\bfp)\,E_P(\bfpp)\, |A_{V_\mu}^{DP}(\bfp,\bfpp)|^2 }
           { B^D(\bfp)\,B^P(\bfpp) }
   },
\end{align}
where we employ the renormalization factor $Z_V$
calculated non-perturbatively in Ref.~\cite{Tomii:2016xiv}.
The relevant semileptonic form factors are then extracted
via Eq.~(\ref{eqn:intro:ME}).

% The $D\!\to\!\pi$ and $D\!\to\!K$ form factors are extracted from
% these $P\!\to\!D$ matrix elements by noting
% $f_+^{DP}(q^2)\!=\!f_+^{PD}(q^2)$ and $f_-^{DP}(q^2)\!=\!-f_-^{PD}(q^2)$.

%// simple analysis ------------------------------------------------------------

\section{Parameter dependence of form factors}\label{sec:simple}

%// param dependence 

\begin{figure}[tb] % no figure before 1st section
  \centering
  \includegraphics[width=0.49\linewidth,clip]{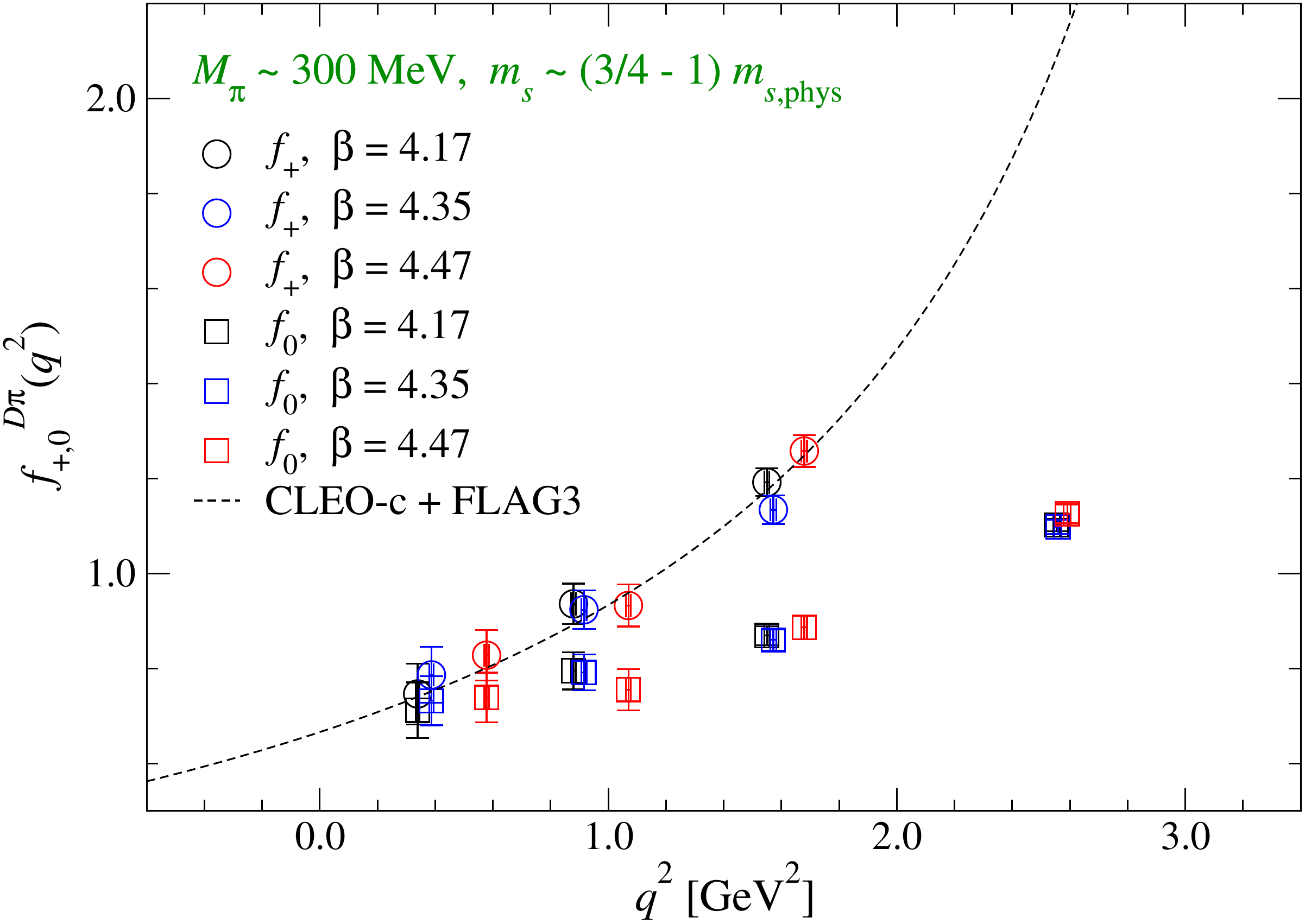}
  \hspace{1mm}
  \includegraphics[width=0.49\linewidth,clip]{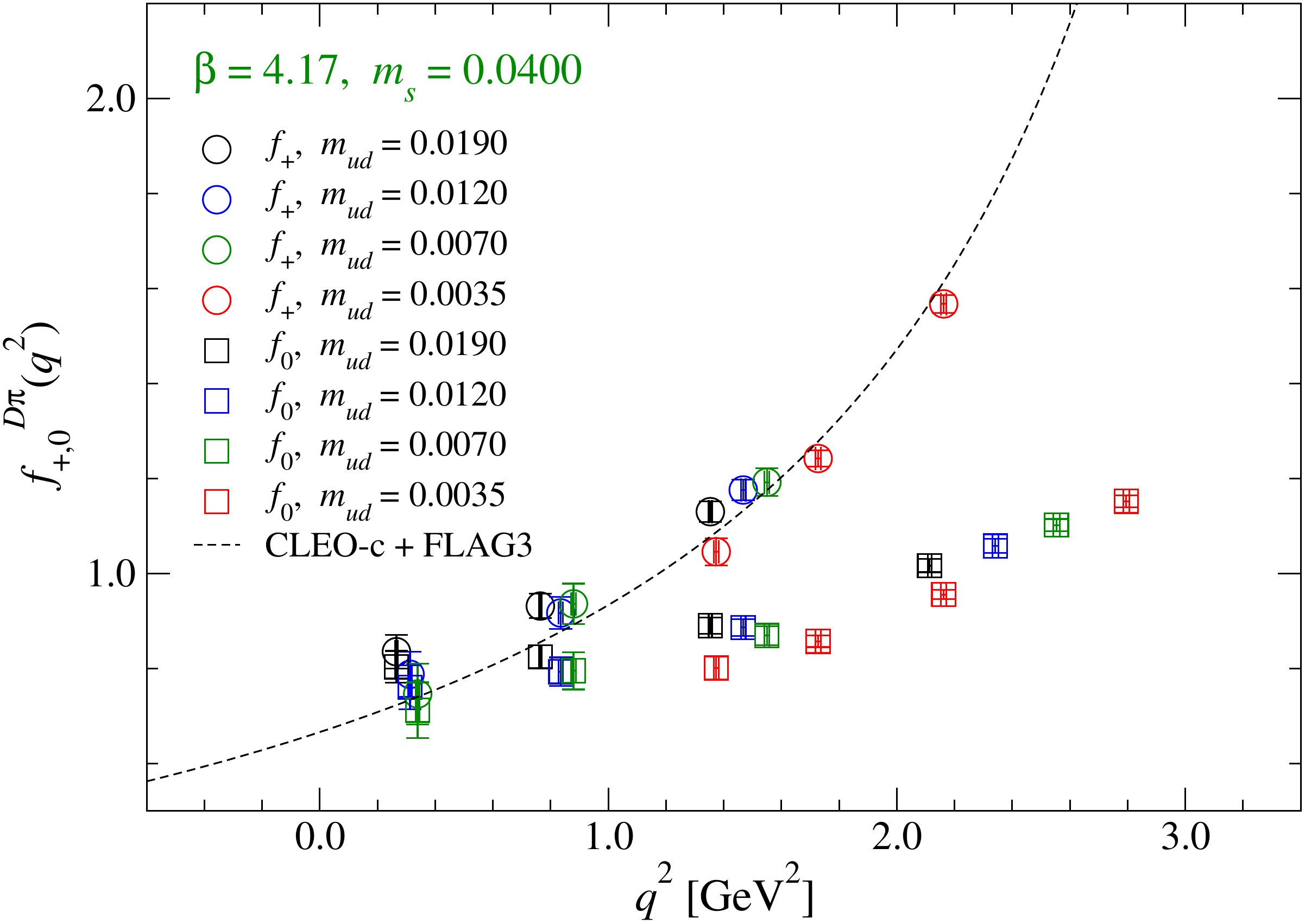}

  \caption{
    Left panel:
    comparison of $D\!\to\!\pi$ form factors among different values
    of $a^{-1}$. We plot data of $\fpDpi(q^2)$ (circles)
    and $\fzDpi(q^2)$ (squares) at $M_\pi\!\sim\!300$~MeV
    and at the larger $m_s$
    as a function of $q^2$.
    Right panel:
    comparison of $D\!\to\!\pi$ form factors among different values of $M_\pi$.
    We plot data at $\beta\!=\!4.17$ and $m_s\!=\!0.0400$.
    In both panels,
    the dashed line shows 
    the Becirevic-Kaidalov parametrization~\cite{Becirevic:1999kt}
    of the CLEO-c data for $\fpDpi$~\cite{Besson:2009uv}
    with the normalization $\fpDpi(0)$ fixed to the world average
    by the Flavor Lattice Averaging Group (FLAG)~\cite{Aoki:2016frl}.
  }
  \label{fig:simple:param-dep}
  \vspace{-3mm}
\end{figure}

Figure~\ref{fig:simple:param-dep}
compares results for $\fpzDpi$ among different values of
$a^{-1}$ (left panel) and $M_\pi$ (right panel).
We observe reasonably good consistency among the lattice data
and also with the experimental data of $\fpDpi$~\cite{Besson:2009uv}.
This suggests that
$\fpzDpi$ mildly depends on $a^{-1}$ and $M_\pi$
from our simulation region down to $a\!=\!0$
and the physical pion mass $M_{\pi,\rm phys}$.
We note that our choice of the lattice action and cutoffs
also leads to small discretization errors
for the decay constant $f_{D_{(s)}}$~\cite{Fahy:2017enl}.

%// z parameter

Similar to our observation in the previous report~\cite{Kaneko:2017sct},
the momentum transfer dependence of $\fpzDP(q^2)$
is well described by the model independent expansion
in terms of a small parameter~\cite{Bourrely:1980gp}
\begin{align}
   z(t,t_0)
   \ = \ 
   \frac{ \sqrt{t_+-q^2} - \sqrt{t_+-t_0} }
        { \sqrt{t_+-q^2} + \sqrt{t_+-t_0} }
   \label{eqn:simple:z}
\end{align}
now including the largest $a^{-1}$ and smallest $M_\pi$.
Here $t_+\!=\!(M_D+M_P)^2$ is the $DP$ threshold energy.
The free parameter $t_0$ is set to an optimized value
$t_0\!=\!(M_D+M_P)(\sqrt{M_D}-\sqrt{M_P})^2$,
such that the magnitude $|z|$ is small in the semileptonic region.
In this article, 
we employ the so-called 
Bourrely-Caprini-Lellouch (BCL) parametrization~\cite{Bourrely:2008za}
\begin{align}
   \fpzDP(q^2)
   \ = \ 
   \frac{1}{B_{\{+,0\}}(q^2)} \sum_{k=0}^{N_{\{+,0\}}} a_{\{+,0\},k}\, z^k
   \label{eqn:simple:bcl}
\end{align}
to determine the normalization $\fpDP(0)\!=\!\fzDP(0)$.

For $\fpDKpi$,
we use the vector meson mass $M_{D_s^*}$ ($M_{D^*}$)
measured at each simulation point
as the resonance mass in the Blaschke factor
$B_+(q^2)\!=\!1-q^2/M_{D_s^*(D^*)}^2$.
The $z$ dependence of $\fpDP$ is well approximated
by the pole factor $1/B_+(q^2)$,
and $N_+\!=\!1$ already gives reasonable $\chi^2/{\rm d.o.f.}\!\lesssim\!1$.
Since we have not yet calculated the scalar meson masses,
$M_{D_{s0}^*}$ nor $M_{D_0^*}$,
it is not clear whether there exist isolated poles below the threshold $t_+$
at the simulated values of $M_\pi$.
We employ a simple linear fit in $z$
by setting $N_0\!=\!1$ and $B_0(q^2)\!=\!1$,
which also yields $\chi^2/{\rm d.o.f.}\!\lesssim\!1$.
The systematic uncertainty due to the choice of the parametrization form
is estimated by testing the quadratic function
($N_+\!=\!2$) for $\fpDP$,
and the Blaschke factor $B_0(q^2)\!=\!1-q^2/M_{D_{s0}^*(D_0^*)}^2$ for $\fzDP$
with the experimental values of $M_{D_{s0}^*}$ and $M_{D_0^*}$.
The systematic error turns out to be comparable or smaller
than the statistical uncertainty.

%// continuum + chiral extrapolation 

\begin{figure}[tb] % no figure before 1st section
  \centering
  \includegraphics[width=0.49\linewidth,clip]{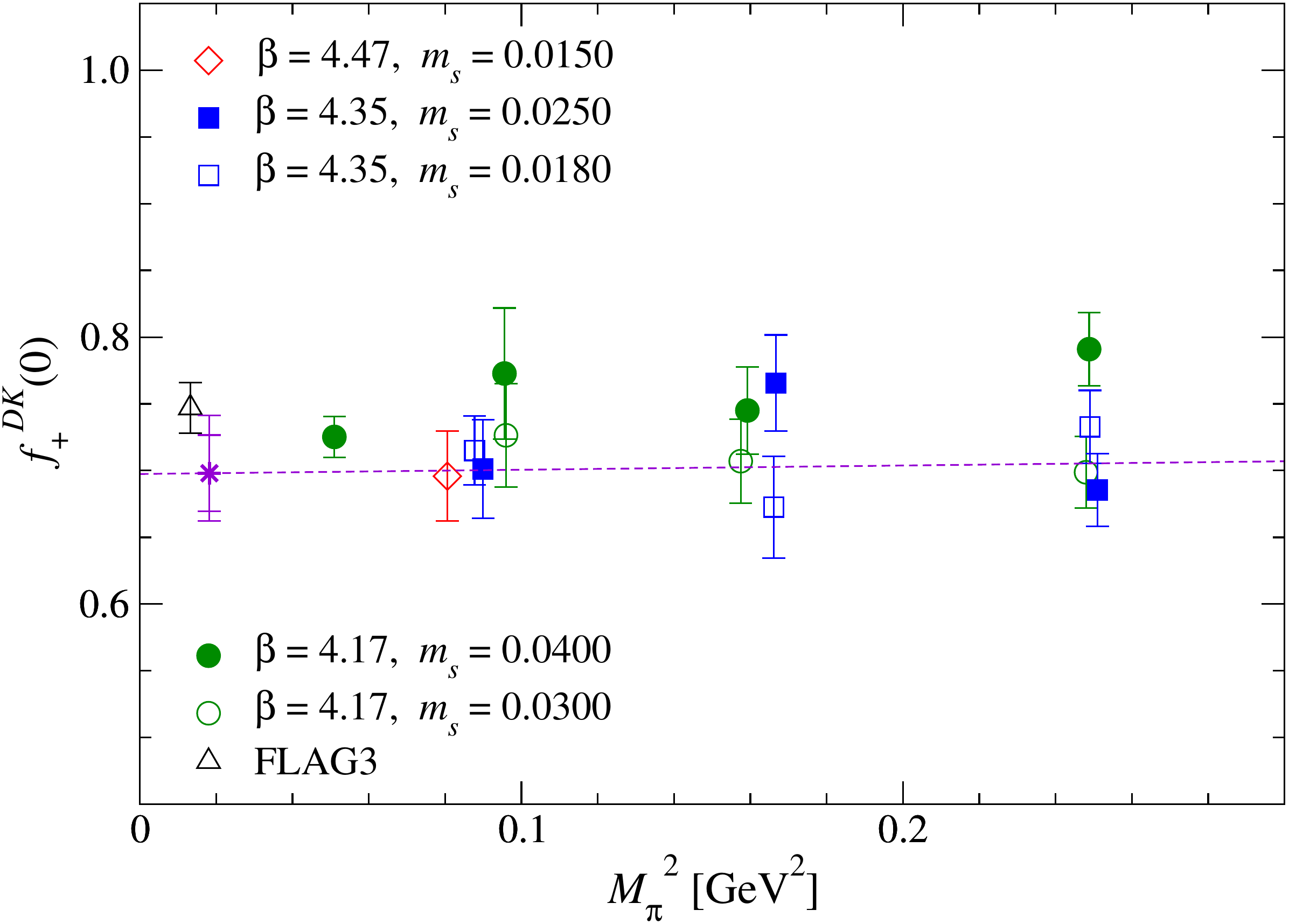}
  \hspace{1mm}
  \includegraphics[width=0.49\linewidth,clip]{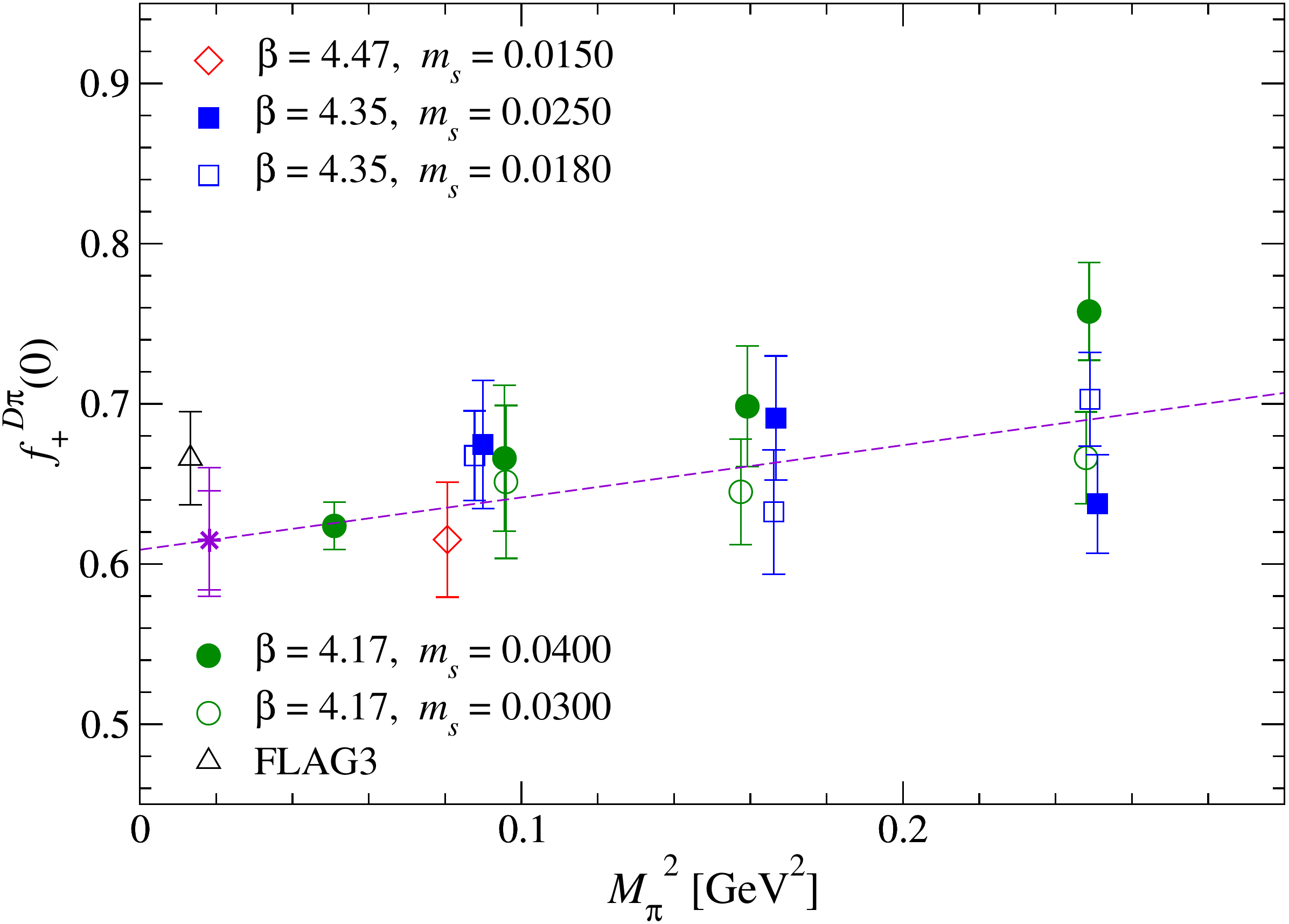}

  \caption{
    Continuum and chiral extrapolation for $\fpDK(0)$ (left panel)
    and $\fpDpi(0)$ (right panel).
    We plot $\fpDKpi$ at different $a$ and $m_s$ by different symbols
    as a function of $M_\pi^2$.
    The dashed line represents the fit line (\ref{eqn:simple:cont+chiral_fit})
    at $a\!=\!0$ and the physical strange quark mass $m_{s,\rm phys}$.
    We also plot the recent FLAG average~\cite{Aoki:2016frl}
    by the open triangle.
  }
  \label{fig:simple:cont+chiral_fit}
  \vspace{-3mm}
\end{figure}

Having observed the mild dependence on the lattice spacing and
quark masses in Fig.~\ref{fig:simple:param-dep},
we extrapolate $\fpDP(0)$
to the physical point in the continuum limit 
by using a simple linear extrapolation 
\begin{align}
   \fpDP(0)
   \ = \ 
   c^{DP}
 + c_{a}^{DP} a^2
 + c_{\pi}^{DP} M_\pi^2
 + c_{\eta_s}^{DP} M_{\eta_s}^2,
   \label{eqn:simple:cont+chiral_fit}
\end{align}
where $M_{\eta_s}^2\!=\!2M_K^2-M_\pi^2\!\propto\!m_s$.
Figure~\ref{fig:simple:cont+chiral_fit} shows
this continuum and chiral extrapolation for $\fpDK$ and $\fpDpi$.
As expected from Fig.~\ref{fig:simple:param-dep},
we obtain good values of $\chi^2/{\rm d.o.f}\!\sim\!1.0$\,--\,1.3,
and most coefficients (except $c_\pi^{D\pi}$) are consistent with zero.
We estimate the systematic uncertainty
by including a quadratic term for the $M_\pi^2$ dependence of $\fpDpi(0)$
and by removing each linear term for other parameter dependences.
Our numerical results are 
\begin{align}
   f_+^{DK}(0)
   \ = \ 
   0.698(29)_{\rm stat}
   (-18)_{q^2\to 0}\left(^{+32}_{-12}\right)_{a\to 0, \rm chiral},
   \\
   f_+^{D\pi}(0)
   \ = \ 
   0.615(31)_{\rm stat}
   \left(^{+17}_{-16}\right)_{q^2\to 0}\left(^{+28}_{-7}\right)_{a\to 0, \rm chiral},
\end{align}
where the first error is statistical error,
and the second (third) error is the systematic uncertainty
due to the $q^2$ (continuum and chiral) extrapolation.
As seen in Fig.~\ref{fig:simple:cont+chiral_fit},
our results are consistent with the recent FLAG averages~\cite{Aoki:2016frl}.

%// CKM elements 

We obtain $|V_{cs}|\!=\!1.035(^{+64}_{-53})_{\rm lat}(5)_{\rm exp}$
and $|V_{cd}|\!=\!0.232(^{+17}_{-13})_{\rm lat}(3)_{\rm exp}$
by using recent experimental input $|V_{cs(d)}|\,\fpDKpi(0)$
compiled by the Heavy Flavor Averaging Group (HFAG)~\cite{Amhis:2016xyh}.
This estimate satisfies CKM unitarity in the second row
$|V_{cd}|^2+|V_{cs}|^2+|V_{cb}|^2\!=\!1.13(0.14)$,
where $|V_{cb}|$ has small effects, and hence 
the tension between the exclusive and inclusive decays does not distort
unitarity.
It is well known that 
the uncertainty of $\fpDKpi(q^2_{\rm ref})$ at a reference $q^2_{\rm ref}$
increases
toward the conventional reference point $q^2_{\rm ref}\!=\!0$\,\footnote{
This is a good reference point for $K\!\to\!\pi\ell\nu$,
since $SU(3)$ flavor breaking effects are quadratic
in small symmetry-breaking parameter $m_s\!-\!m_u$~\cite{Ademollo:1964sr}.}.
As in recent lattice studies of $B$ meson semileptonic decays,
it is better to make comparison between lattice QCD and experiments
in a wide region of $q^2$.

%// global fit -----------------------------------------------------------------

\section{Global fit of form factors}\label{sec:global}

%// parameterization 

\begin{figure}[tb] % no figure before 1st section
  \centering
  \includegraphics[width=0.49\linewidth,clip]{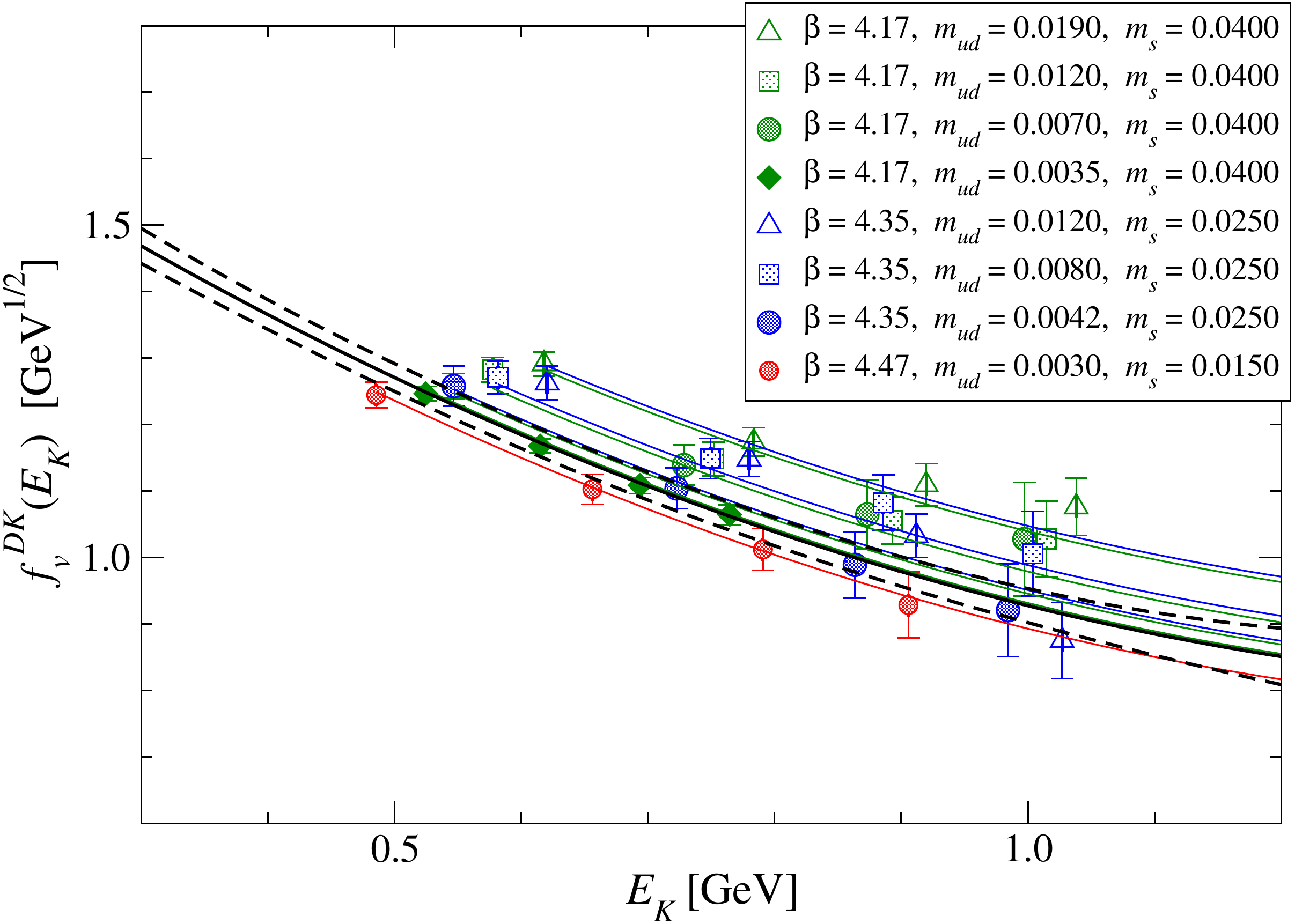}
  \hspace{1mm}
  \includegraphics[width=0.49\linewidth,clip]{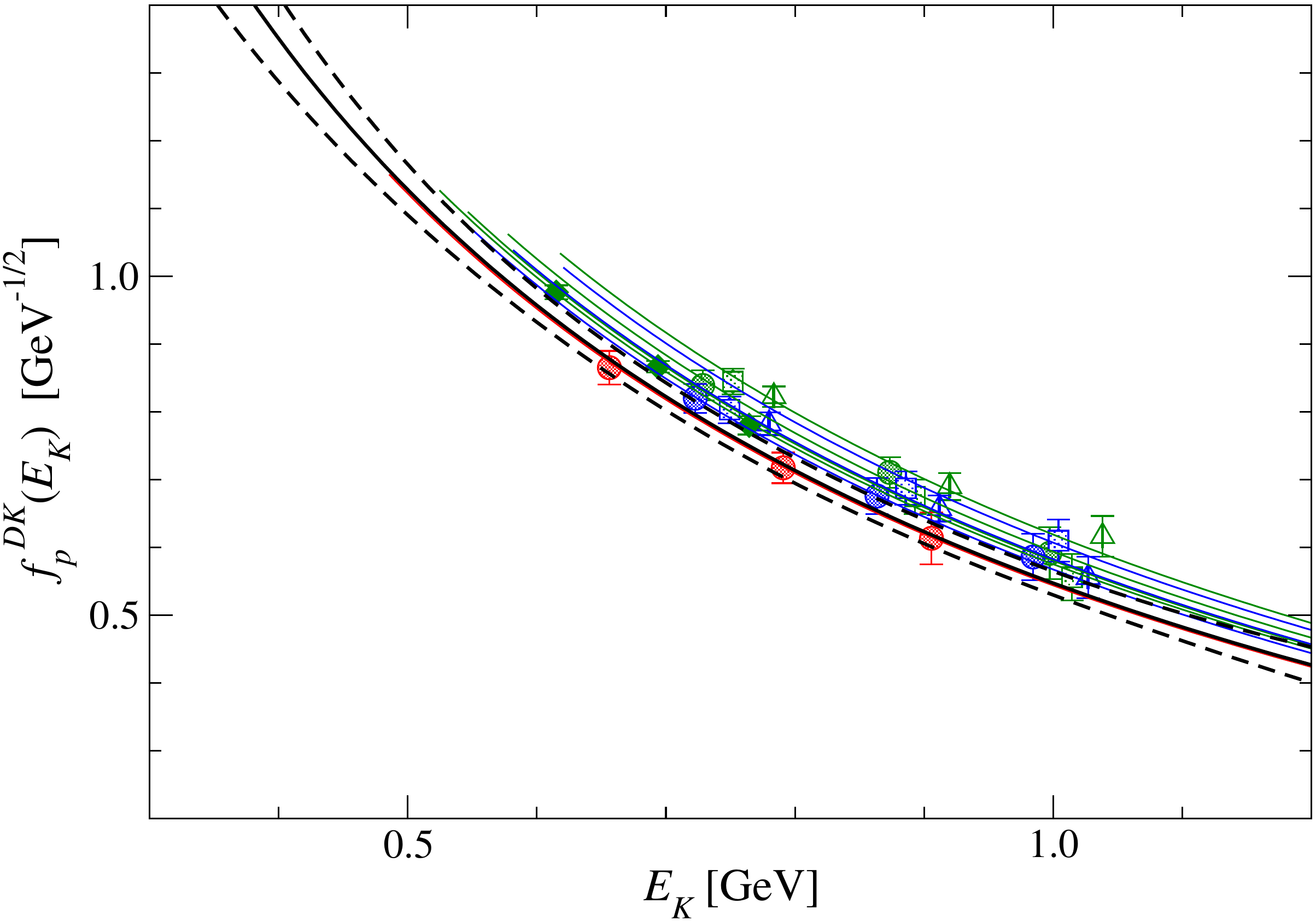}

  \caption{
    Global fit for $\fparaDK$ (left panel) and $\fperpDK$ (right panel).
    For clarity, 
    we plot $\fhqetDK$ only at larger $m_s$ as a function of $E_K$,
    while the global fit is carried out by using all data.
    Different symbols show data at different $a$ and $M_\pi$.
    The thick lines show the fit curve at $M_{\pi,\rm phys}$ and at $a\!=\!0$.
  }
  \label{fig:global:global_fhqetDK}
  \vspace{-3mm}
\end{figure}

In this article, 
we carry out a global fit of the form factors
as a function of $a$, $M_\pi^2$ and $M_{\eta_s}^2$
based on SU(2) hard pion heavy meson chiral perturbation theory
(hard pion HMChPT)~\cite{Bijnens:2010jg},
while the pionic logarithms have been calculated
in both heavy meson and relativistic formulations.
To this end, we use the form factors in the following decomposition
convenient for the heavy meson formulation
\begin{align}
   \langle P(\pp) | V_\mu | D(p) \rangle
   \ = \ 
   \sqrt{2M_D}
   \left\{
      v_\mu\,\fparaDP(E_P) + \left( \pp - E_P v\right)_\mu\,\fperpDP(E_P)
   \right\},
   \label{eqn:global:ME_HQET}      
\end{align}
where $v\!=\!p/M_D$ is the $D$ meson 4-velocity,
and $E_P\!=\!v\pp$ is the light meson energy in the $D$ rest frame.
We note that
$1/m_Q$ dependence of $\fhqetDpi$ turns out to be mild 
even at $m_Q\!=\!m_c$ in our study of the $B\!\to\!\pi\ell\nu$ decay~\cite{brian}.
We leave a global fit based on the relativistic parametrization~(\ref{eqn:intro:ME}) and that in conventional HMChPT~\cite{Falk:1993fr,Becirevic:2003ad}
for our future analysis to study the systematic uncertainty.

%// parameterization

We employ a parametrization form 
\begin{align}
   \fparaDP(E_P)
   \ = \ 
   c_v^{DP}
   \left\{
      f_{\rm log}^{DP}
    + c_{v,\pi}^{DP} M_\pi^2 + c_{v,\eta_s}^{DP} M_{\eta_s}^2
    + c_{v,E}^{DP} E_P      + d_{v,E}^{DP} E_P^2     
    + c_{v,a}^{DP} a^2
   \right\}, 
   \label{eqn:global:fpara}
   \hspace{10mm}
   \\ 
   \fperpDP(E_P)
   \ = \ 
   \frac{c_p^{DP}}{E_P+\Delta M_D}
   \left\{
      f_{\rm log}^{DP}  
    + c_{p,\pi}^{DP} M_\pi^2 + c_{p,\eta_s}^{DP} M_{\eta_s}^2
    + c_{p,E}^{DP} E_P      + d_{p,E}^{DP} E_P^2     
    + c_{p,a}^{DP} a^2
   \right\}
   \label{eqn:global:fperp}
\end{align}
from our observation in the previous subsection.
Namely, we include the linear terms to describe
the mild dependence on $a$, $M_\pi^2$ and $M_{\eta_s}^2$,
whereas a quadratic term is added for the dependence on $E_P$,
which is related to $q^2$ through $E_P\!=\!(M_D^2+M_P^2-q^2)/2M_D$.
We include the pole factor %1/(E_P+\Delta M_D)$
with $\Delta M_D\!=\!M_{D_{(s)}^*}-M_D$ for $\fperpDP$,
which is proportional to the matrix elements
with the spatial vector current $\langle P | V_k | D \rangle$.
The chiral logarithm is given as
$f_{\rm log}^{DP}\!=\!(c_{\rm log}^{DP}/\Lambda_\chi^2)M_\pi^2 {\rm log}\left[ M_\pi^2/ \Lambda_\chi^2\right]$
with $\Lambda_\chi\!=\!4 \pi f$,
$c_{\rm log}^{D\pi}\!=\!-3(1+3g^2)/4$ and
$c_{\rm log}^{DK}\!=\!1/2$.
We fix the decay constant in the chiral limit $f$
to the FLAG average~\cite{Aoki:2016frl}
and the $D^*D\pi$ coupling to $g\!=\!0.61$~\cite{Lubicz:2017syv}.

%// global fit for HQET FFs

As shown in Fig.~\ref{fig:global:global_fhqetDK},
the parametrizations (\ref{eqn:global:fpara}) and (\ref{eqn:global:fperp})
describe our data well
and yield $\chi^2/{\rm d.o.f}\!\sim\!0.9$ $(\fhqetDK)$
and 1.2 $(\fhqetDpi)$.
The coefficients for the $a$, $M_\pi^2$ and $M_{\eta_s}^2$ dependences
are not large. % as in the case with $\fpzDP$.
We emphasize that $c_{\{v,p\},a}^{DP}$ is consistent with zero,
which suggests small discretization error for $\fhqetDP$.

%// global fit for HQET FFs

\begin{figure}[t] % no figure before 1st section
  \centering
  \includegraphics[width=0.49\linewidth,clip]{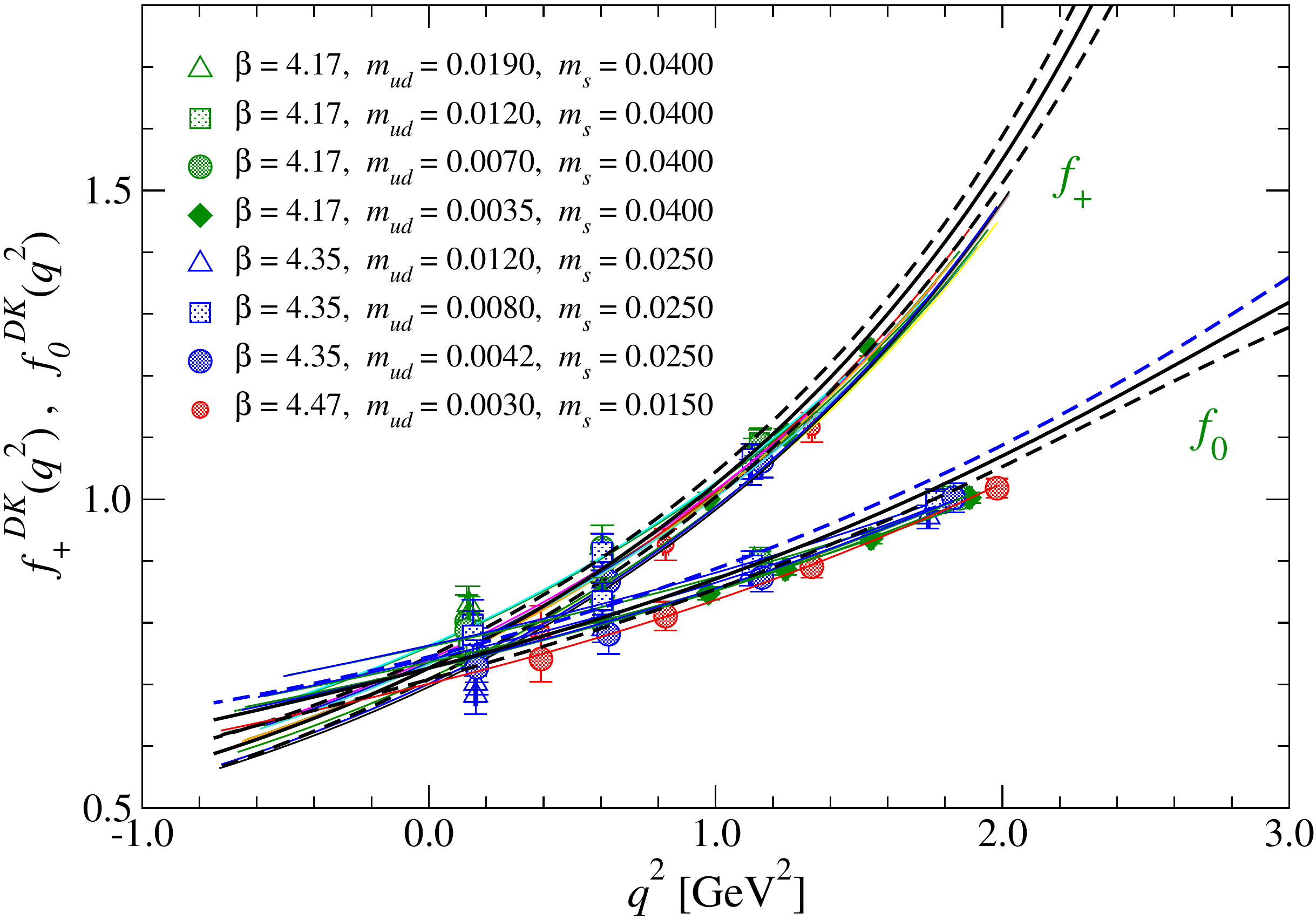}
  \hspace{1mm}
  \includegraphics[width=0.49\linewidth,clip]{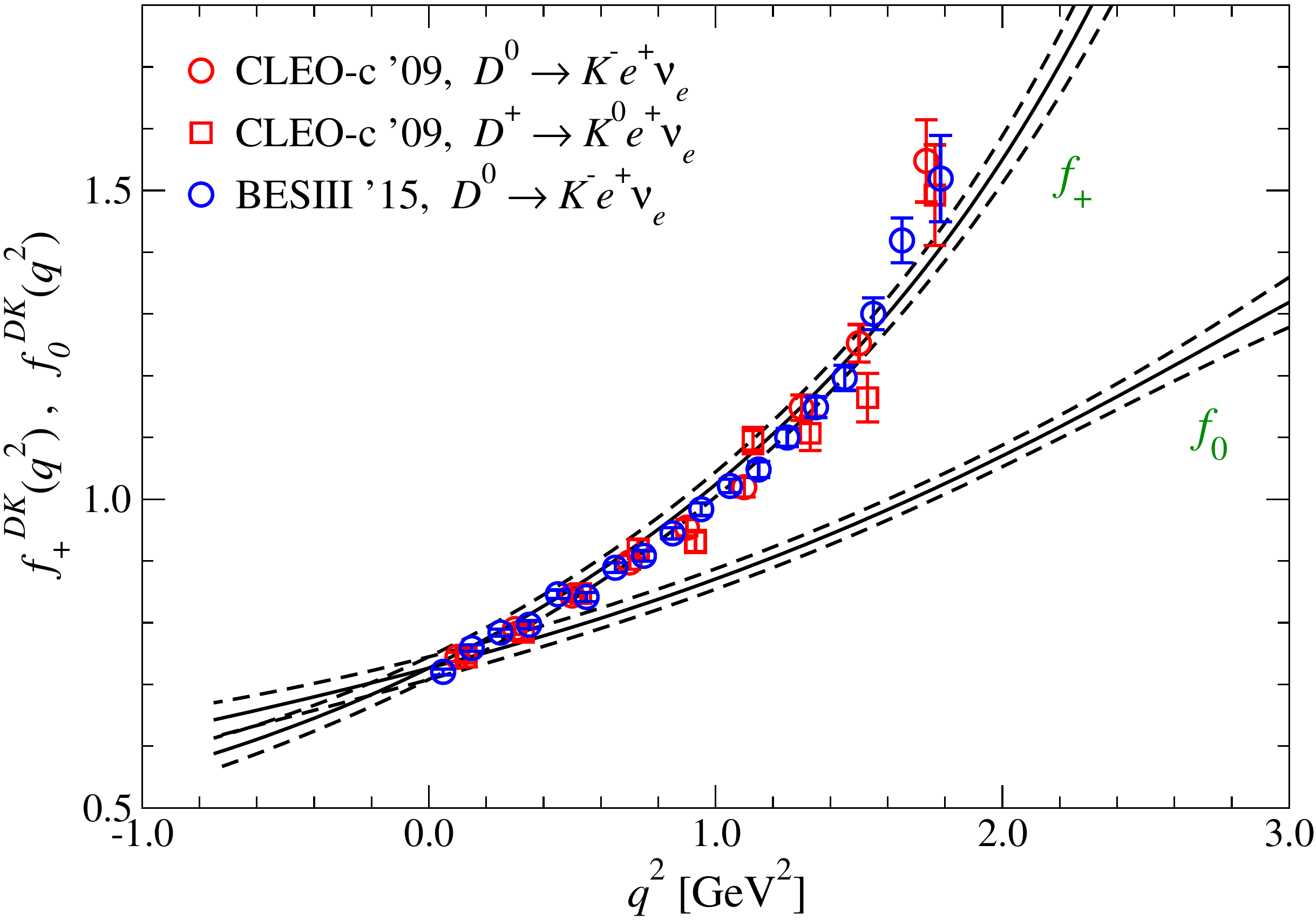}

  \caption{
    Left panel:
    global fit for $\fpzDK$ reproduced from that for $\fhqetDK$.
    For better visibility, we plot data at larger $m_s$
    as a function of $q^2$.
    The green, blue and red symbols show data
    at $a^{-1}\!\sim\!2.5$, 3.6 and 4.5~GeV, respectively,
    whereas the different symbols show data at different $M_\pi$ values.
    The thick lines show the fit curve at the physical point and $a\!=\!0$.
    We also plot the fit curves at simulation points by thin lines,
    which are however indistinguishable at the scale of this figure
    due to the mild $a$ and $M_\pi$ dependence of $\fpzDP$.
    Right panel:
    comparison of form factor shape between lattice (thick solid lines)
    and experimental (circles and squares) data.
  }
  \label{fig:global:global_fpzDK}
  \vspace{-3mm}
\end{figure}

The global fit for $\fhqetDP$ can be translated into that for $\fpzDP$
by using the kinematical relation
\begin{align}
   \fpDP(q^2)
   \ = \ 
   \frac{1}{\sqrt{2M_D}}
   \left\{ \fparaDP(E_P) + \left(M_D-E_P\right)\,\fperpDP(E_P) \right\},
   \hspace{19mm}
   \\
   \fzDP(q^2)
   \ = \ 
   \frac{\sqrt{2M_D}}{M_D^2-M_P^2}
   \left\{
      \left( M_D - E_P \right) \fparaDP(E_P)
    + \left( E_P^2 - M_P^2 \right) \fperpDP(E_P)
   \right\}.
\end{align}
The left panel of Fig.~\ref{fig:global:global_fpzDK} shows
the global fit for $\fpzDK$,
which reproduces the data at simulation points reasonably well.

%// comparison with experiment

From experimental data of the partial decay rate $\Delta \Gamma_i$
together with the CKM matrix element $|V_{cs(d)}|$,
we can estimate the vector form factor $\fpDKpi$ as
\begin{align}
   \fpDKpi(q_i^2)
   \ = \ 
   \frac{1}{|V_{cs(d)}|}
   \sqrt{
     \frac{24\pi^3}{G_F^2} \frac{1}{p_i^3} \frac{\Delta \Gamma_i}{\Delta q^2_i}
   },
\end{align}
where $G_F$ is the Fermi constant,
$\Delta q^2_i$ is the size of the $i$-th $q^2$ bin,
and $p_i$ is the light meson momentum in the $D$ rest frame for the $i$-th bin.
Note that experimental data are available for the light lepton modes
$D\!\to\!P\{e,\mu\}\nu$, and hence have low sensitively to $\fzDP$,
which is suppressed by the lepton mass squared $m_l^2$.
We estimate $\fpDP$ from the CLEO-c~\cite{Besson:2009uv}
and BESIII~\cite{Ablikim:2015ixa} data of $\Delta \Gamma_i$
and the HFAG values of the CKM matrix elements~\cite{Amhis:2016xyh}.
The right panel of Fig.~\ref{fig:global:global_fpzDK}
confirms good agreement in the form factor shape between
our lattice data and experimental data.
This also suggests that we obtain the CKM matrix elements
close to their HFAG values when we determine them
as a relative normalization factor between the lattice and experimental data.

%// conclusion -----------------------------------------------------------------
\section{Conclusions}\label{sec:concl}

In this article, we update the status of our study of
the $D$ meson semileptonic decays.
Form factors are calculated on fine lattices with cutoffs up to 4.5~GeV
by using the M\"obius domain-wall action for both light and charm quarks.
We observe good consistency of the normalization $\fpDKpi(0)$
with previous lattice
calculations and the form factor shape is nicely consistent
with experiment.

Having observed small discretization errors at $m_c$,
it is important to extend our simulations to the $B$ meson decays.
At this conference,
we have reported our studies of the $B\!\to\!\pi\ell\nu$~\cite{brian}
and inclusive~\cite{inclusive} decays.
Another interesting future direction is an extension to
form factors to quantify new physics contributions.
Our analysis of the $D$ meson tensor form factors are underway.

%// acknowledgement ------------------------------------------------------------

\section*{Acknowledgments}
% \vspace{5mm}

Numerical simulations are performed on the IBM System Blue Gene Solution at KEK
under its Large Scale Simulation Program (No.~16/17-14)
and on the Oakforest-PACS supercomputer operated by the Joint Center for
Advanced High Performance Computing (JCAHPC).
% through the HPCI System Research Projects (Project ID: hp170106).
%
This research is supported in part by JSPS KAKENHI Grant Number 
JP26247043 and JP26400259,
and by MEXT as “Priority Issue on post-K computer”
(Elucidation of the Fundamental Laws and Evolution of the Universe)
and the Joint Institute for Computational Fundamental Science.

\bibliography{lattice2017}

%%%%%%%%%%%%%%%%%%%%%%%%%%%%%%%%%%%%%%%%%%%%%%%%%%%%%%%%%%%%%%%%%%%%%%%%%%%%%
\end{document}